\documentclass[12pt]{article}
\usepackage{graphicx}
\usepackage{aasms4}

\def\lapprox{\hbox{\lower .8ex\hbox{$\,\buildrel < \over\sim\,$}}}
\def\gapprox{\hbox{\lower .8ex\hbox{$\,\buildrel > \over\sim\,$}}}

\begin{document}

 \title{The light curve and  distance 
of Kepler supernova: news from four centuries ago}

 \author{Pilar Ruiz--Lapuente$^{1,2}$}

 \altaffiltext{1}{Instituto de F\'{\i}sica Fundamental, Consejo Superior de 
 Investigaciones Cient\'{\i}ficas, c/. Serrano 121, E--28006, Madrid, Spain}
 \altaffiltext{2}{Institut de Ci\`encies del Cosmos (UB--IEEC),  c/. Mart\'{\i}
 i Franqu\'es 1, E--08028, Barcelona, Spain} 

 \begin{abstract}

\noindent 
We study the light curve of SN 1604 using the historical data collected
 at the time of observation of the outburst. Comparing the 
 supernova with recent SNe Ia of various rates of decline after maximum light, 
 we find that 
 this event looks like a normal SNIa (stretch $s$ close to 0.9: 0.9 $\pm$ 0.13),
 a fact which is also
 favoured by the late light curve. The supernova is heavily obscured by 2.7$\pm$
 0.1 magnitudes in V. We obtain an estimate of the distance to
 the explosion with a value of $d = 5 \pm 0.7$ kpc. 
 This can help to settle 
 ongoing discussions on the distance to the supernova. It also shows
that this supernova is of the same kind as those of the SNIa sets
that we use for cosmology nowadays.

 \end{abstract}

 \keywords{Supernovae, general; supernovae, Type Ia; supernovae, individual, 
SN 1604; cosmic distance scale}

 \section{Introduction}

 \noindent
 The supernova of 1604 was observed by Johannes Kepler and other
 astronomers in Europe, Korea, and China. Not so many years before, the
 supernova SN1572, currently named Tycho's supernova, was the subject of 
  extensive studies, both astronomical and philosophical,
 by Tycho Brahe
 (Brahe 1603). 

 \bigskip

 \noindent
 The supernova in 1604 inspired as well observational
 measurements and philosophical considerations by Kepler.
 The philosophical disquisitions were on the nature of the heavens. 
 The idea of  crystal spheres
   carrying each one a planet, and rolling over one another,
 had already been shattered by the  discovery of SN 1572. 
 The possibility of a heaven that gives birth to natural objects
 grew along the following years, and was  reenhanced by the discovery
  of the supernova of 1604. Kepler wrote his reflections  
 in a book called {\it De Stella Nova in Pede  Serpentariis} (1606). 
 The collection of his observational records on SN 1604 and those 
 of colleagues are found mostly in his compendium  {\it Gesammelte Werke}
(collection of Kepler's works made in 1938).

 \bigskip

 \noindent
 Because of the importance of Kepler as a physicist, and of
 the records collected 
 by him from SN 1604, this supernova is often called Kepler's supernova.

 \bigskip

 \noindent
 For many years, it was not clear which type of supernova it was. There 
 were discussions on whether it was a
 core--collapse supernova or a thermonuclear one. 

 \bigskip

 \noindent
 X--ray studies have shown that the  O/Fe ratio in the remnant 
 of SN 1604 corresponds 
 to that of a Type Ia supernova (Reynolds et al. 2007). In this supernova 
 remnant, there are signs that the system might have been relatively massive,
  one of its components creating a detached circumstellar shell (CSM) of 
 1 M$_{\sun}$ expanding into the interstellar  medium (Vink 2008). Recent 
 studies of this CMS suggest that it had lost contact with the inner evolving 
 stars years before the explosion (Katsuda et al. 2015). 

 \bigskip

 \noindent
 In agreement with the abovementioned, some authors have suggested that 
 the companion star could have been an Asymptotic Giant Branch 
 (AGB) star that had lost its
 envelope (Chiotellis et al. 2012). 

 \bigskip

 \noindent
 All the discussions concerning the progenitor of Kepler supernova are
 haunted by the lack of precision in the estimated distance. The supernova 
 is located well above the Galactic plane, at l$=$ 4.5$^{o}$, b$=$6.4$^{o}$ 
 (Vink 2008). It has an angular size of 2 arcmin.
 The  suggested distances to the remnant lies between 3 and 7 kpc. One of
 the first distances were obtained by Reynoso and Goss (1999), using the 
 H I absorption to the remnant to give a constraint of 4.8 $<$ $d_{Kepler}$ $<$ 
 6.2 kpc. The first distance given by Sankrit et al (2005), using the 
 proper motion of the optical filaments, had a value of 3.9$_{-0.9}^{+1.4}$ kpc. 
 Vink (2008) gave 
 a distance $\sim$ 6 kpc, from arguments related to the velocity of the 
 forward shock. Based on the non--detection of TeV gamma--rays by HESS, 
 Aharonian et al. (2008) suggested a distance of $>$ 6 kpc. 
 Most recently, Sankrit et al. (2016) have revisited the method based  on the
  proper motion of the filaments to derive as distance 5.1$^{+0.8}_{-0.7}$ kpc.   

 \bigskip

 \noindent
 In the present work, we go back to the method, previously used 
 for understanding Tycho supernova in nowadays cosmological terms 
 (Ruiz--Lapuente 2004), and we make use of the historical records 
 of SN 1604 to reconstruct its light curve. We determine the best 
 stretch of the light curve and the distance to the remnant.

\bigskip

\noindent
Knowing the distance to the supernova is crucial in the searches 
from a possible surviving companion in this supernova (Kerzendorf et al. 2014;
Ruiz--Lapuente  2014). This knowledge allows to place luminosity 
limits to the searches and inform with accuracy of the kind of potential
companions surveyed. 

 \section{Observations}

 \noindent
 We have compared the historical records 
 gathered by European and Korean astronomers at the time of the
 explosion in 1604  with the family of SNe Ia as 
 known today. The supernova occurred in a region of the sky  which was often 
 observed, because the supernova appeared 3 degrees to the northwest of 
  Mars and Jupiter, which were in conjunction, and about 4 degrees to the 
 east of Saturn.  So, there were plenty of observations that  allowed
 to place the time of the appearence of the supernova at October 10, being 
 unseen before. The early light curve is very complete, with daily 
 reports on the brightness of the supernova. The discovery preceded by
 20 days the visual maximum. This means that we have data for three weeks,
 a lapse which compares well with the best follow--ups currently done on early  
 discovered supernovae. But, unlike Tycho's SN which was circumpolar, 
 SN 1604 was observed from Europe and Korea at relative low latitude,  
and during the
months of November and December, the supernova 
was not observable during the night. 

 \bigskip

 \noindent
 The most recent reconstruction of the historical records has been made
 by Clark and Stephenson (1977), who compare the records obtained
 by Korean astronomers with those obtained by the European astronomers,
 previously examined by Baade (1943).
 The comparison between the Korean and European records 
 gives a good agreement in the early part of the light
 curve, before and around maximum,  with some disagreement in the phase after 
 maximum.

 \bigskip

\noindent
 The high surveillance of the supernova in epochs before its maximum
(the visual   maximum was on Nov 1, according to the fit to the
light curve) was done in Europe by a group of anonymous observers. Only
on Oct 17 Kepler had a chance to see the supernova. We assign 0.5 mag
to the individual magnitude errors in these premaximum observation
as they were mostly done by untrained observers.

\bigskip

\noindent
Concerning the records after maximum, they are often single records 
and sometimes by untrained observers. Tycho Brahe's observations
of the supernova of 1572 showed that the Danish astronomer achieved 
the maximum accuracy of the human eye and was able to distinguish 
changes in brightness at a few tenths of a magnitude. Therefore, we could
assign 0.25 mag and even 0.2 mag to some of his observations. 

\bigskip

\noindent
Johannes Kepler, the magnificient Imperial Mathematician who followed
Tycho Brahe in ths court of Prague,  would bring physics to a 
modern era, when the laws of motion of the planets discovered by 
him were explained by Isaac Newton. He took interest in the supernova
SN 1604, but his contribution can not be compared with his explanation
of the orbit of Mars.  Kepler used glasses as he was shortsighted .
He had difficulties
in differentiating the brightnesses differing by 0.25 magnitudes or more.
CS77 found the comment by Kepler in {\it De Stella Nova} that ``the
star it was seen with almost the same magnitude during the whole 
month of October''. Kepler started to see the ``nova'' on Oct 17, 
but until Nov 1st it had an increase in magnitude by more than 0.3 mag.
Kepler also wrote on Feb 6  ``I left the observatory, not sure whether
I had seen any trace of the star. Therefore, it seems to have become 
too small to be seen even in this clear morning, if it has survived''
(see CS77, p.199).  The other records are from untrained astronomers.
It would have benefitted the reconstruction of the light curve of SN 1604
that the observations by Fabricius had been preserved, as he was known to 
be an accurate astronomer. Baade (1943) was only able to recover some
mention to these observations in Kepler's collected works. Thus, in general,
we judge that the error in the European observations after maximum can be
of 0.5 mag.

\bigskip

\noindent
The descriptions of the brightness of SN 1604 written by the Korean astronomers
are simple and the colors that they gave at premaximum are at odds
with those of the European astronomers. Thus, we assign 0.5 mag uncertainty
 to the
values derived for the brightness, the same error bar as for the premaximum 
values by European observers.

\bigskip

\noindent
 From both the early and the total light curve, 
 it can be excluded that Kepler SNIa were a subluminous SN 1991bg--like event. 
 The  
 European records have a late slope declining more slowly
 than a SN 1991bg-like SNIa. 
 However, it does not look either as an overluminous SN 1991T--like one. 
 The best agreement
 would be with something  of the type of a ``normal'' SNIa (see 
 Figures 1 and 2 ).

 \subsection{Early light curve up  to 60 days past maximum}

\bigskip

\noindent
SN Ia are not standard candles but calibrated ones. There is 
a well--known correlation between the brightness at peak of the
light curve and its rate of decline. Phillips (1993) gave a
first correlation in terms of a parameter $\Delta m_{15}$, which is
the number of magnitudes of decline of the B light curve in 15 days 
after maximum brightness. Hamuy et al. (1996) used it to calibrate 
the Calan Tololo SN Ia. This way to obtain the 
absolute magnitude of a SNIa in relation to the Hubble constant 
did also include a correction for the extinction suffered by the
supernova due to dust, mostly in the host galaxy.

\bigskip

\noindent
The early way to calibrate SNe Ia treated separately 
the correction from stretch (this one including the intrinsic color
of a given SNIa of a particular stretch) and the extinction by dust.
Later on, Tripp (1998) advocated for the use of two simultaneous determinations
of the parameter of stretch and the one of color, the latter 
taking into account the intrinsic color of the SNIa and the extrinsic 
one due to dust. 

\bigskip

\noindent
In our present case, the extinction by dust in the Galaxy is very large and
it is very well known. This is why we prefer to use the early version
of stretch that did not require to fit a global color term, but to estimate 
the extinction. The excess E(B-V) 
is then estimated separately, for SN 1604.

\bigskip
 
\noindent
 Therefore, the data on SN 1604 are compared using the stretch
  factor ${\it  s}$  for the 
 characterization of the rate of decline (Perlmutter et al. 1999; 
 Goldhaber et al. 2001; Nobili et al. 2003). The stretch factor $s$ 
 method, used by the Supernova Cosmology Project,  quantifies the
 decline rate of the supernova from data extending  up to 60 days 
 after maximum. In the absence of a mesurement of the brightness at 
 maximum, the method allows  to locate the event within
 the family of light curves of SNe Ia.

 \bigskip

 \noindent
 The best agreement of the light curve of SN 1604
 is with a supernova of $s \sim$ 0.9 
($s=0.9 \pm 0.13$).
 In this sense, as in the case of SN 1572, we show a comparison in 
 Figure 1 with the normal supernova SN 1996X, which has an $s$ of 0.889.
 Also for comparison, we show the template light curve of a SN Ia with 
$s = 0.62$, a subluminous 91bg-type,  and the template of a $s = 1.2$ 
SN Ia like the overluminous SN 1991T.
 This Figure is centered on the early light curve. 

 \bigskip

 \noindent
 Comparing the historical records with a stretch $s = 0.9$ supernova template, 
 we obtain a $\chi^{2} = 13.235$ for 13 degrees of freedom, that  gives  a 
  reduced $\chi^{2}/d.o.f$ of 1.02, which is a good fit. In contrast, the
 fit to the template of a fast--declining, underluminous SNIa like SN 1991bg, 
 of stretch $ s= 0.62$, has a $\chi^{2}$ of 21.65 for 7 degrees of freedom, that
 is a $\chi^{2}/d.o.f$ of 3.09. In the other extreme, the fit to the template of
 a slowly--declining, overluminous SNIa, like SN 1991T ($ s = 1.2$) is the 
worst  one, 
 with a  $\chi^{2}$ of 60.30 for 13 degrees of freedom, which gives a reduced 
  $\chi^{2}/d.o.f$ of 4.64.

 \begin{figure}
 \centering
 \includegraphics[width=1.0\columnwidth]{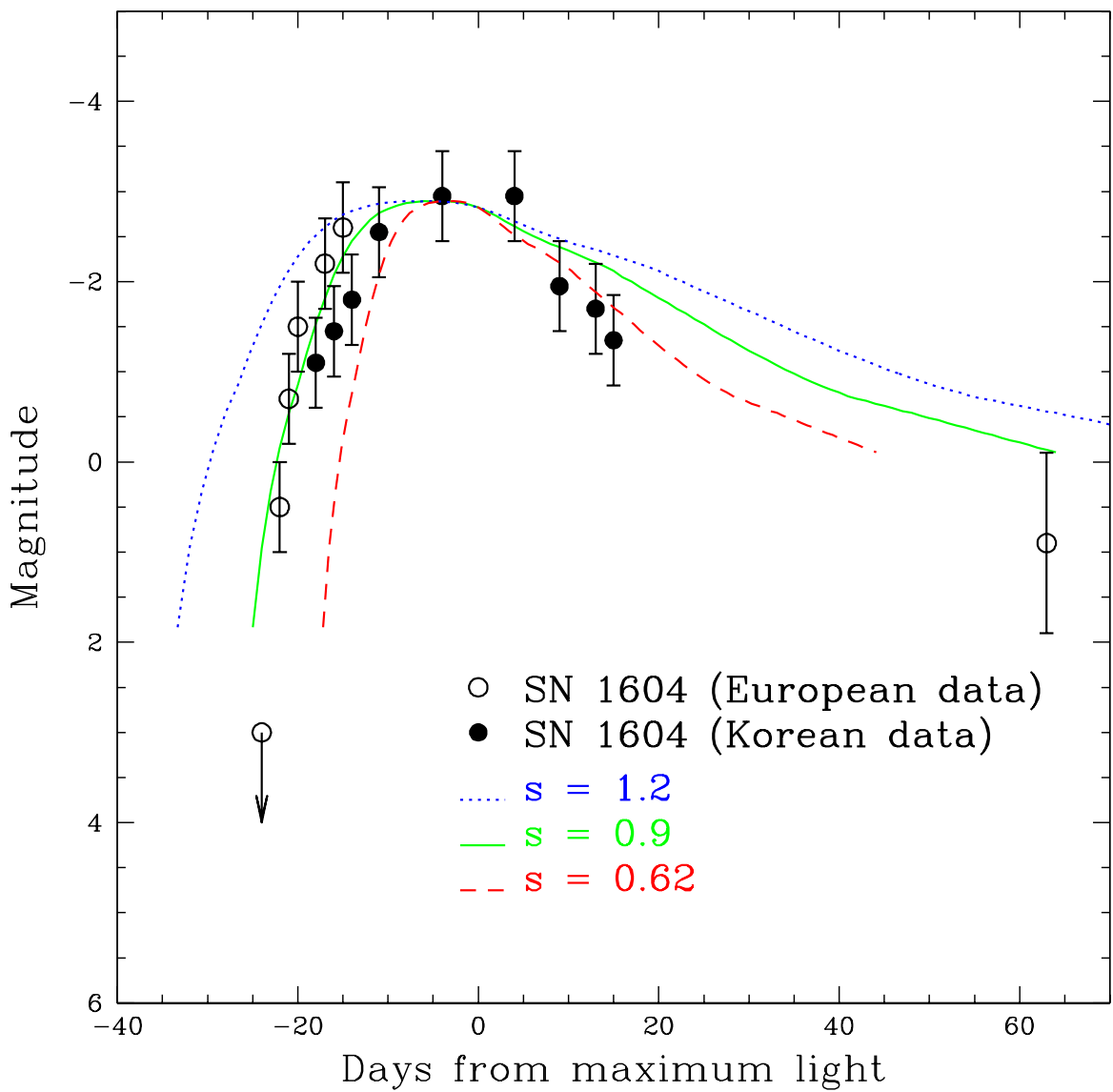}
 \caption{The visual light curve of SN 1604 around maximum light is shown 
 here. The Korean records are indicated by filled circles and the European ones
 by open circles. The observations are compaired with a normal
 SNIa with stretch $s = 0.9$, with an overlumious and slow--declining $s = 1.2$
 SNIa; and  with an underluminous and fast declining $s = 0.62$ SNIa. }
 \label{Figure 1}
 \end{figure}

 \subsection{Late light curve}

 \noindent
 If one compares the magnitudes attributed by Baade and those attributed
by Clark and Stephenson (1977) along the post--maximum decline, 
one often finds that Baade (1943) substracts 0.2 magnitudes to what would
have been the right magnitude assignment. 
So, the European magnitudes in CS77
  are  a bit different from those of  
 Baade (1943), and brighter.  Examples of this are the records 
of April and August.
 When the recorded comparison says that the star is as bright as $\eta$ Oph, 
 which has a visual magnitude of 2.43, Baade assigns
 to the record 2.6. CS77
  assign 2.40 in one ocassion (April 21), but in another one they judge 
the magnitude to be 2.25 (April 12). When the last  written records 
 say that the star is as bright as $\xi$ Oph, which is
 4.39 magnitudes in V, Baade (1943) rounds the number to 4.5, and
 Clark and Stephenson (1977) give 4.45.
 The two last records are similar in Baade (1943)
 and in CS77, with only a 0.1 magnitude difference in the assigned brightness
 of the supernova.  Baade (1943) assigns to the record ``fainter than 
 $\xi$ Oph'', 5 and 4.8 in visual magnitude, whereas CS77 assign  4.95 and 4.7. 
Therefore, the disagreement is not significant. However it shows a lack 
 of complete agreement and it suggests that we should assign an error of 0.5 
magnitudes to the records. 

\bigskip

\noindent
CS77 located a few postmaximum data in the light curves reported 
by the Korean astronomers. They plot huge error bars for those records,
which come from the mean of several observations in every case. These are
the only points for which CS77 have reported error bars. We assign errors
of 0.7 magnitudes to these data.

 \bigskip

 \noindent
  In Figure 2, we show the supernova light curve compared with those of 
SN 1996X, SN 1991T and
 SN 1991bg.  The European records at postmaximum are decisive in classifying 
SN 1604 as a normal SNIa, specially the last four. A comparison of the light
 curve of SN 1991bg with that of SN 1604 gives a  $\chi^{2}$122.48 for 21
d.o.f, this is $\chi^{2}/d.o.f$ 5.8. It is not acceptable. If we want o test
the similarity with SN 1991T, we get $\chi^{2}$ $=$ 87.83 for 31 d.o.f, so 
$\chi^{2}/d.o.f$ $=$ 2.83.  
 A comparison of the 
light curve o SN 1996X  and SN 1604 gives a $\chi^{2}$ 40.507 for 30 degrees
of freedom, so $\chi^{2}/d.o.f = $ 1.35, which seems acceptable.

 \bigskip

 \noindent
 After 100 days, the rate of decline is 1.37 $\pm$ 0.12 V magnitudes 
 in 100 days, according to Baade (1943). This also places the light curve 
 decline amongs those of 
 normal SNe Ia, which have decline rates of 1.35--1.5 magnitudes in 100
 days (see declines for 1990N, SN 1999bu; for references, Ruiz--Lapuente 2004).

 \begin{figure}
 \centering
 \includegraphics[width=1.0\columnwidth]{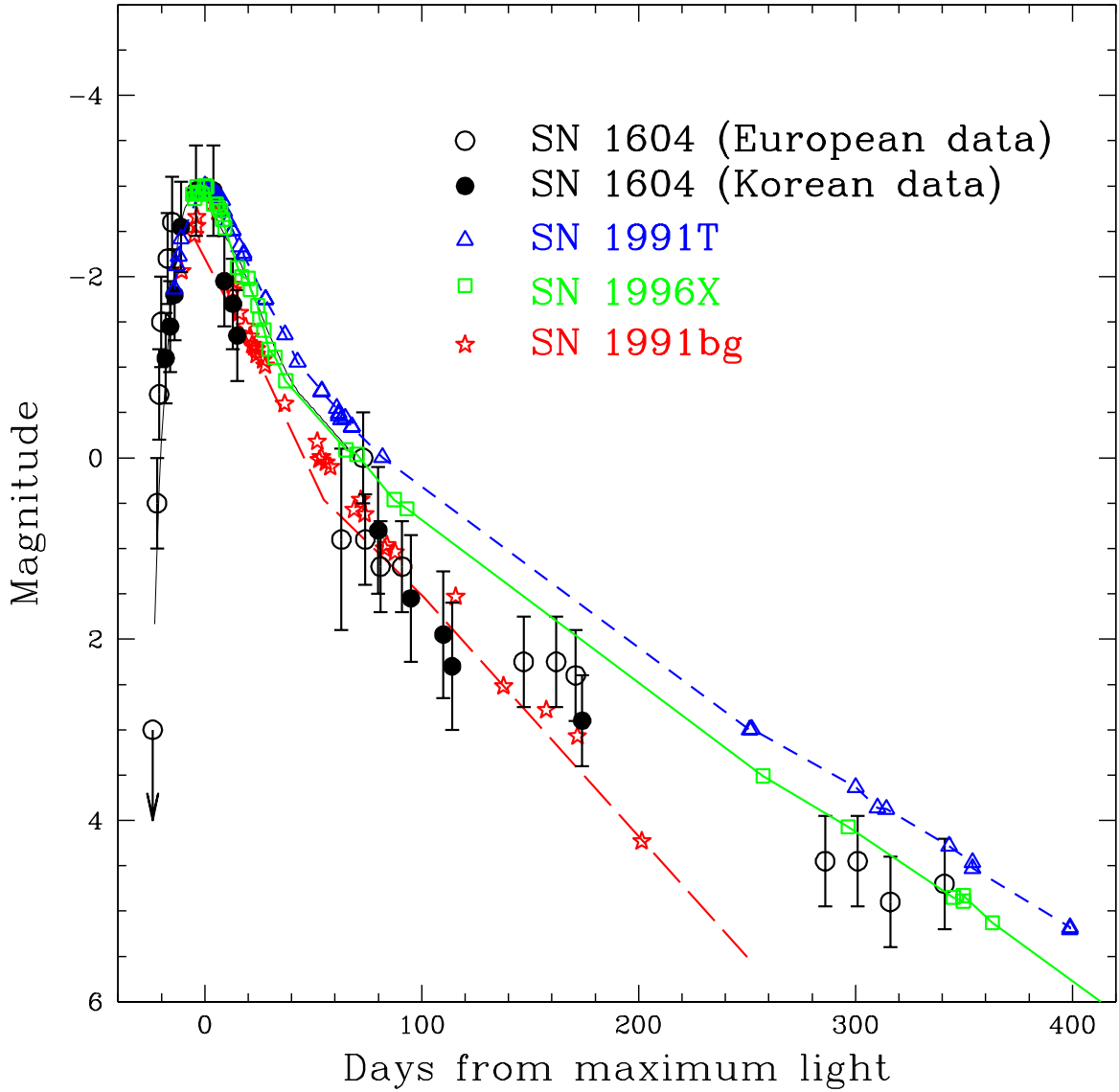}
 \caption{Visual light curve of SN 1604 from the records collected
by Baade (1943) and Clark and Stephenson (1977). The supernova is 
compared with the normal supernova SN 1996X, as well as with the overluminous 
SN 1991T and the subluminous SN 1991bg}
 \label{Figure 2}
 \end{figure}

 \subsection{Kepler was neither a SNIa-CMS nor any other  type of peculiar SNIa}

 \noindent
 SNe Ia that interact heavily with the circumstellar medium have been 
 named SNe Ia--CMS. They are charactherized by a narrow H$\alpha$ line
 on top of an overluminous spectrum. They have much more luminous and flat
  declines than 91T--like SNe Ia (Silverman et al. 2013).

\bigskip

\noindent
  SNe Ia--CMS (only 0.1--1 $\%$ of all SNe Ia) 
 were discovered by Hamuy et al (2003),
 in SN 2002ic, which suggested that this supernova could have arised from a binary system containing a C+O white dwarf plus a massive (3--7 M$_{\sun}$) AGB star, where the total mass loss in H can reach a few solar masses. Another well discussed SNIa--CMS is PTF11kx, observed by Dilday et al. (2012), who suggested
that it came 
 from a symbiotic nova progenitor. Soker et al. (2013) noted, however, that 
the mass
 around PTF11kx is too high to have been produced by a recurrent nova,
 and suggested
 a violent prompt merger of a white dwarf with the core of a massive AGB star.

 \bigskip

 \noindent
 Katsuda et al. (2015) find that Kepler SN, unlike SNIa--CMS, should have a 
 dense/knotty CMS located far away from the progenitor star. They discuss 
 that the CMS knots were already $\sim$ 2 pc away from the progenitor star
 at the time of the SN explosion. They also add as  evidence that the
 interactions between the CMS and the blast wave started $\geq$ 200 yrs after
 the explosion, And the third piece of evidence is the difference between 
 the historical light curve of Kepler and the light curves of SNIa-CMS. 

\bigskip

 \noindent
 In the paper of Katsuda et al.(2015), it is argued, from the
  amount of Fe found in
 the X--ray spectrum of the supernova, that it should have syhthesized 
 around 1 M$_{\sun}$ of Ni. We wonder whether the Fe found could not come
 from some source different from 
 $^{56}$Ni decay, since the light curve does not seem to
 follow the path of an overluminous  SNIa. Patnaude  et al. (2012)
 show that some models of SNe Ia with an energy output of 1.4 10$^{51}$
 ergs and having their ejecta interacting with a constant density 
 ambient medium can explain the X--ray emission without making 
 Kepler supernova overluminous. 

 \bigskip

\noindent
Kepler supernova is clearly not one of the classes of peculiar SNe Ia: 
SNIax, which fall below  the Phillips relation being significantly fainter
than the bulk of SNe Ia; neither are those fast declining SNe Ia that unlike
SNe Iax are not faint at maximum (see examples and references in
Ruiz--Lapuente 2014). On the other extreme, it is not a Super--Chandrasekhar
supernova with an overluminous magnitude and slow rate of decline.
Definitevely, SN 1604 lies in the bulk of cosmological SNe Ia.

 \section{The distance to Kepler supernova}

\bigskip

 \noindent
 We aim now to estimate the distance to Kepler. We think that we have
 the elements to provide a fairly good estimate.

 \bigskip

 \noindent
 First of all,  we have the new measurements of extinction in the Galaxy
 provided by Schlafly \&  Finkbeiner (2011), based on comparison of predicted
colour of stars with observed ones in various surveys. These authors 
 find that the previous
 estimates by Schlegel, Finkbeiner \& Davis  (2009) overestimated extinction 
 values in  the Milky Way by 14 $\%$. The previous values used the infrared
emission in the cosmic microwave background (CMB)  radiation foregrounds 
to extract a map of extinction in the Galaxy.
The updated value of extinction in the
 direction of Kepler supernova by Schlafly \& Finkbeiner (2011) is 
 A$_{V}$=2.7.

 \bigskip

 \noindent
 Blair et al 1991 estimated from the knots in SN 1604 
 where H$_{\beta}$ could not
 be measured accurately but  H$_{\alpha}$ was detected, an excess E(B-V)=0.9. 
 This estimate agrees well with the fact that the supernova at the early premaximum  should have had an intrinsic B-V = --0.05 while it had 
 an excess E(B--V) = 0.9 from the historical records.  Those two estimates
 indicate an extinction
 A$_V$$=$ 3.1 $\times$ E(B-V) $=$ 2.7$\pm$ 0.1.
 We thus assume A$_V=$2.7 $\pm$ 0.1 for SN 1604,

 \bigskip

 \noindent
 Concerning the absolute maximum in the visual of Kepler supernova, it should 
 be $M_V$ $\sim$ --19.2 magnitudes. We have already shown that the light curve 
 of Kepler supernova gives a better fit for a normal SN Ia than for an 
 overluminous or an underluminous one. We can apply the absolute calibration
 for the maximum light of SN Ia derived by Hamuy et al. (1996) using
 the Calan Tololo SN Ia sample. This calibration is 
 $M_{V_{max}} +  5\ log\ (H_{0} / 65) = -19.26 \pm 0.12$. 

 \bigskip

 \noindent
 If we take 67.8 $\pm$ 0.9 km s$^{-1}$ Mpc$^{-1}$ as the value of $H_{0}$, then
 the small error in $H_{0}$ carries a $\Delta M_{V_{max}}$ $=$ 0.03 .
 Then we take into account the error associated to the strech fit.
 It is carried out taking into account the uncertainty in 
fitting the early maximum and the factor multiplying this rate--of--decline 
 fit. This gives
 a final $\Delta M_{V}$ $\sim$ 0.2 mag.  A M$_{V}$$=$ -19.2  is 
 actually in agreement with a 
 previous indication obtained by modeling the late phases of normal
 SN Ia (Ruiz--Lapuente 1997).
 The brightness of Kepler SNIa at maximum was -3  mag, as bright as 
 Saturn. 
 Then, replacing the values in
 $M_{V} = m_{V} -5\ log\ d_{p} + 5 - A_{V}$,
 we obtain $d = 5 \pm 0.7$ kpc. 
 The error in the distance corresponds to a 
 0.2 mag error in the estimated peak magnitude in V plus 0.1 in extinction.

 \section{Summary and conclusions}

\noindent
The nature of SN 1604, the most recently observed Galactic SN, was an object 
of debate since its discovery, Johannes Kepler being its most famous observer
and the most prominent figure in the ensuing disquisitions. Even when its 
SN nature was acknowledged, there were still discussions about its 
classification, either as Type I or a Type II SN. This point of debate was 
settled some 10 years ago, by X--ray observations of the remnant. In spite of
being acknowledged as a Type Ia, thermonuclear SN, its further classification 
within the SN Ia family has remained unclear.

\noindent
 We have used the historical records on SN 1604, coming from European and
Korean astronomers, to reconstruct the light curve of this SN Ia. That has 
been based on the combination of the attribution of magnitudes by Baade (1945) 
and by Clark \& Stevenson (1977) to the ancient records. We assign a
 precision according to the information about the records.

\noindent
The data have then been fitted with template light curves, parameterized by 
the stretch factor $s$, The best fit corresponds
 to a ``normal'' SN Ia, 
($s=$ 0.9 $\pm$ 0.13). 
The fit excludes both overluminous events like SN 1991T and
subluminous ones like SN 1991bg. 

\noindent
The absolute magnitude of SN 1604 at maximum
should have been $M_{Vmax} = -19.2$ and the error is calculated according
 to the error in magnitude coming from  the error in stretch and in the
calibration linked with H$_{0}$. It is of 0.2 magnitudes in V.

\noindent
SN 1604 was heavily obscured. Therefore, the amount of extinction suffered 
plays an important role in the determination of its distance. Based on 
several coincident measurements, we have $A_{V} = 2.7 \pm 0.1$ mag.  

\noindent
We obtain a distance to SN 1604 of $d = 5\pm0.7$ kpc, in agreement with 
recent estimates based on the proper motions of the optical filaments of the 
remnant of the SN, but discarding suggested distances of $\sim$ 6 kpc or more.
That is very important for the direct search of a possible surviving companion
to the SN.

\bigskip

\noindent
{\bf Acknowledgements}

\noindent
I would like to thank discussions on Kepler's SNIa with Keiichi Maeda at the
MIAPP on ``The Physics of Supernovae'' during the summer of 2016. I am also
grateful to Rahman Amanullah for making possible a comparison with various
versions of the stretch correction. This research was supported by the
 Munich Institute for Astro- and Particle Phyisics (MIAPP) of the DFG cluster
 of excellence ``Origin and Structure of the 
Universe''. The author is supported by AYA2015--67854-P from the Ministry 
of Industry, Science and Innovation and the FEDER funds.

\newpage

 \begin{table}
         \centering
         \caption{Reduction of Korean  estimates for
                  SN 1604 to magnitudes}
         \label{tab:brightness_k}
         \begin{tabular}{lllllc} 
                 \hline
 Date &    & Phase   & V mag  & Error & Observer \\
      &    & Adopted & Adopted &      & Reference\\
 \hline
 1604 & October 14 &  -17 & -1.1  & 0.5    &     Korean astronomers \\
      & October 16 &  -15 & -1.45 & 0.5    &     -----              \\
      & October 18 &  -13 & -1.8  & 0.5    &     -----              \\
      & October 19 &  -12 & -2.55 & 0.5   &      -----              \\
      & October 28 &   -3 & -2.95 & 0.5   &      -----              \\
      & November 5 &   +4 & -2.95 & 0.5    &     -----              \\  
      & November 10 &  +9 & -1.95 & 0.5    &     -----              \\
      & November 14&  +13 & -1.7  & 0.5    &     -----              \\
      & November 16&  +15 & -1.35 & 0.5    &     -----              \\
 1605 & January  20&  +80 & +0.8  (mean)& 0.7 &   -----             \\
      & February 4 &  +95 & +1.55 (mean)& 0.7 &  -----              \\
      & February 19& +110 & +1.95 (mean)& 0.7 & -----               \\
      & February 23& +114 & +2.3  (mean)& 0.7 & -----               \\
      & April 24   & +174 & +2.9  & 0.5 &   -----                   \\
                 \hline
                 \hline
      \end{tabular}
  \end{table}

\newpage

 \begin{table}
         \caption{Reduction of European estimates for
                  SN 1604 to magnitudes}

         \label{tab:brightness_e}
         \begin{tabular}{llllll} 
 \hline
 Date &    & Phase   & V mag   & Error & Observer \\
      &    & Adopted & Adopted &       & Reference\\
 \hline
 1604 & October  8& -23 & +3 or fainter& 0.5 &  Several  \\
      & October  9& -22 & +0.9 & 0.5 & Anonymous physician and others \\
      & October 10& -21 & +0.5 & 0.5  & Anonymous physician and others \\
      & October 11& -20 & -0.7 & 0.5 & Anonymous physician and others \\
      & October 12& -19 & -1.5 & 0.5 & Roeslin \\
      & October 15& -16 & -2.2 & 0.5 & Physician, Fabricius, Maestlin$^{2}$ \\
      & October 17& -14 & -2.6 & 0.5 & Kepler$^{1}$ \\
 1605 & January 3& +63 & +0.9 (mean)& 1.0 & various and Kepler$^{1}$\\
      & January 13& +73& 0.00 & 0.5 &  Kepler$^{1}$  \\
      & January 14& +74& +0.9 & 0.5 & Fabricius$^{2}$ \\ 
      & January 21& +81& +1.2 & 0.5 & Maestlin$^{2}$ \\
      & January 31& +91& +1.2 & 0.5 &   \\ 
      & March 28 &+147&  2.25  & 0.5 &  \\
      & April 12 &+162& +2.25& 0.5 &Fabricius$^{2}$  \\
      & April 21 &+171& +2.40& 0.5 &Kepler$^{1}$ \\\
      & August 12--14&284--86& +4.45 & 0.5&Kepler$^{1}$ \\ 
      & August 29  &+301& +4.45 & 0.5 &Kepler$^{1}$ \\
      & September 13&+316& +4.95 & 0.5 &Kepler$^{1}$ \\\
      & October 8  &+341& +4.7 & 0.5 &Kepler$^{1}$ \\
 \hline
  \hline
 \end{tabular}

$^{1}$ Kepler J., Gesammelte Werke (ed. Caspar 1938), 1, 161-164

$^{2}$ Kepler J., Opera Omnia (ed. Frish 1859)  
\end{table}

\newpage

 \begin{table}
     \caption{Colors of Kepler's Supernova}

       \label{tab:color}
  \begin{tabular}{llllll}
 \hline
 \hline
 Date  &       & Phase   & Description & B - V   & Observer \\ 
       &       & Adopted &             & Adopted & Reference\\
 \hline
 1604 & Oct. 8 & -23        & Not seen    &         & Several  \\
      & Oct. 9 & -22        & Like Mars   & 1.36    & Several  \\
      & Oct. 15& -16     & Like Jupiter& 0.82    & Unknown physician\\
      & Oct. 16&         &             &         & Fabricius and others$^{1}$\\
 \hline
 \hline
  \end{tabular}

$^{1}$ Kepler J., Opera Omnia (ed. Frish 1859)
\end{table}

\end{document}